\def\etal{{\it et al.\thinspace}}
\def\mearth{{\rm\,M_\oplus}}

\documentclass[12pt,preprint,psfig]{aastex}

\received{}
\accepted{}
\journalid{}{}
\articleid{}{} 

\lefthead{}
\righthead{}

\begin{document}

\title{Predicting Planets in Known Extra-Solar Planetary Systems III:
Forming Terrestrial Planets}

\author{Sean N. Raymond\altaffilmark{1,2}, Rory Barnes\altaffilmark{1,3}, 
\& Nathan A. Kaib\altaffilmark{1}}

\altaffiltext{1}{Department of Astronomy, University of Washington, Seattle, 
WA, 98195}
\altaffiltext{2}{Current address: Laboratory for Atmospheric and Space
Physics, University of Colorado, Boulder, 80309-0590
(raymond@lasp.colorado.edu)}
\altaffiltext{3}{Current address: Lunar and Planetary Laboratory, University of
 Arizona, Tucson, AZ 85721 (rory@lpl.arizona.edu)}


\begin{abstract}
Recent results have shown that many of the known extrasolar planetary systems
contain regions which are stable for both Earth-mass and Saturn-mass planets.
Here we simulate the formation of terrestrial planets in four planetary
systems -- 55 Cancri, HD 38529, HD 37124, and HD 74156 -- under the assumption
that these systems of giant planets are complete and that their orbits are
well-determined.  Assuming the giant planets formed and migrated quickly, then
terrestrial planets may form from a second generation of planetesimals.  In
each case, Moon- to Mars-sized planetary embryos are placed in between the
giant planets and evolved for 100 Myr.  We find that planets form relatively
easily in 55 Cnc, with masses up to 0.6 Earth masses and in some cases
substantial water contents and orbits in the habitable zone.  HD 38529 is
likely to support an asteroid belt but no terrestrial planets of significant
mass.  No terrestrial planets form in HD 37124 and HD 74156, although in some
cases 1-2 lone embryos survive for 100 Myr.  If migration occurred later,
depleting the planetesimal disk, then massive terrestrial planets are unlikely
to form in any of these systems.
\end{abstract}

\keywords{astrobiology --- planets and satellites: formation --- methods: n-body simulations}

\section{Introduction}

Most planets detected to date around main sequence stars are thought to be
Jovian (gaseous) in nature.  This is known from their large masses, most of
which are larger than 30 $\mearth$ (although smaller planets have been
detected -- e.g. Rivera \etal, 2005), and from transit measurements of the
size of HD209458b to be 1.27 Jupiter radii (Charbonneau \etal 2000).  The
radial velocity technique, which is sensitive to the reflex motion of a
planet's parent star, is unlikely to ever be able to detect Earth-mass planets
in the habitable zones of their parent stars.  The sensitivity of current
surveys is 3-10 $m\,s^{-1}$ (Butler \etal 1996; Baranne et al. 1996; Marcy \&
Butler 1998), while the reflex velocity of the Sun due to the Earth is only
about 9 $cm \, s^{-1}$.  This signal is not likely to be detected by radial
velocity surveys in the near future.  ESA's COROT and NASA's Kepler missions,
to be launched in 2006 and 2007, respectively, hope to be the first to find
Earth-like planets around other stars by looking for transits.  NASA's
Terrestrial Planet Finder (TPF) and ESA's Darwin missions hope to
spectroscopically characterize terrestrial planets around main sequence stars.

Recent results have shown that several of the known planetary systems
contain regions in between the giant planets in which massless test
particles remain on stable orbits for long periods of time.  Barnes \&
Raymond (2004; hereafter Paper I) mapped out these stable regions in
semimajor axis $a$ and eccentricity $e$ space for HD 37124, HD 38529,
HD 74156 and 55 Cnc.  These regions have been mapped in $a$ space
(assuming circular orbits) for $\upsilon$ And (Rivera \& Lissauer
2000), GJ876 (Rivera \& Lissauer 2001) and 55 Cnc (Rivera
\& Haghighipour 2003).

Menou \& Tabachnik (2003) examined the possibility of Earth-sized
planets residing in the habitable zones of known extrasolar planetary
systems (including single planet systems), again using massless test
particles.  They find that roughly one fourth of the known systems can
support a planet in the habitable zone of its parent star, as defined
by Kasting, Whitmire \& Reynolds (1993).

Raymond \& Barnes (2005; hereafter Paper II) tested the stability of
Saturn-mass planets in the regions of four planetary systems in which
test particles had been shown in Paper I to be stable: HD 37124, HD
38529, 55 Cnc and HD 74156.  They found that for Saturn-mass planets,
the stable regions identified in Paper I shrank to a small fraction of
the test particle stable region.

Barnes \& Quinn (2004) tested the stability of seven known planetary
systems, and found that several are on the edge of stability: a small
change in orbital elements can lead to a catastrophic disruption of
the system.  This idea led to the ``packed planetary systems'' (PPS)
hypothesis first suggested by Laskar (1996), and presented in Paper I.
The PPS model suggests that all planetary systems contain as many
planets as they can support without becoming unstable, and implies
that if a stable region exists within a planetary system, then it
should contain an additional planet.  The systems studied in Paper II
are not on the edge of stability, and therefore have enough
``dynamical space'' to harbor additional unseen planets.

Papers I and II dealt solely with the dynamic stability of
hypothetical additional planets in planetary systems.  In this paper
we examine the formation process of terrestrial planets in such a
system.  Giant planets close to their parent stars (e.g., ``hot
jupiters'') are thought to form farther out in the protoplanetary disk
and migrate inward via torques with the gas disk (e.g., Lin \etal
1996).  In order for a terrestrial planet to co-exist with a close-in
giant planet, the terrestrial planet must either (i) form quickly and
survive the inward gas giant migration, or (ii) form from material
remaining after the giant planet has migrated through.

The probability of a planet surviving in the terrestrial region in
scenario (i) is small.  Mandell \& Sigurdsson (2003) showed that in
some cases terrestrial planets can survive the migration of a
Jupiter-like planet through the terrestrial zone.  The fraction of
planets which survive such a migration is a function of the rate of
migration (faster migration implies higher survival rate), and ranges
between 15\% and 40\%.  However, only a small fraction (7-16\%) of the
surviving planets end up with orbits in the habitable zone, meaning
that only 1-4\% of terrestrial planets in the habitable zone are
likely to remain on similar orbits after a migration event.  A much
more likely outcome is that the planet is scattered onto a highly
eccentric orbit with a large semimajor axis (Fig. 3 from Mandell \&
Sigurdsson, 2003).

Can terrestrial planets form from local material after a giant planet
migrates through?  Armitage (2003) showed that in many cases the
post-migration disk of planetesimals is depleted beyond repair.
However, if giant planet migration occurs quickly and early enough in
the disk's lifetime, then enough time remains for a second generation
of planetesimals to form (Armitage, 2003).  Raymond, Quinn, \& Lunine
(2005a) argued that terrestrial accretion can therefore occur in the
presence of one or more close-in giant planets via scenario (ii).
Indeed, several recent results have shown that giant planets may form
in less than 1 Myr via either the bottom-up, core-accretion scenario
(Rice \& Armitage 2003; Alibert, Mordasini \& Benz, 2004; Hubickyj
\etal 2005) or the top-down, fragmentation scenario (Boss 1997; Mayer
\etal, 2002).  Migration begins immediately after (or even during)
formation (Lufkin \etal, 2004) and takes $\sim 10^5$ years or less for
planet larger than 0.1 Jupiter masses (D'Angelo, Kley \& Henning,
2003, and references therein).

Raymond \etal (2005a) show that potentially habitable, terrestrial
planets can form in the presence of a close-in giant planet, assuming
a substantial disk remains reforms after migration.  The orbit and
composition of these planets are strongly affected by the position of
the giant planet.  Hot/warm jupiters at larger orbital radii (up to
0.5 AU) cause the terrestrial planets to be iron-poor and in some
cases drier than for closer-in giant planets, and may reduce the
chances of habitable planet formation.  

Here we simulate the final stages in terrestrial planet formation from disks
of planetary embryos in four known systems: HD 37124 (Butler \etal, 2003),
HD 38539 (Fischer \etal, 2003), 55 Cnc (Marcy \etal, 2002), and HD 74156 (Naef
\etal, 2004), the same four systems we examined in Paper II.  We assume that
embryos form via oligarchic growth (e.g., Kokubo \& Ida 2000), and
allow these bodies to accrete under their mutual gravity and the
gravity of the known planets for 100-200 Myr.  We explore systems in
which rapid, early migration has occurred, thus leaving behind a
substantial planetesimal disk (Armitage 2003).  We also look at cases
in which migration has occurred late in the disk lifetime, leaving
behind a disk with little mass in planetesimals.  In addition, we make
simple comparisons with previous simulations (e.g., Raymond \etal
2004).  In $\S$2 we describe our numerical method and initial
conditions.  We present the results of our simulations in $\S$3, and
conclude in $\S$4.

\section{Method}

\subsection{Initial Conditions}

As discussed above, our formation scenario assumes that the gas giants formed
and migrated through the terrestrial region.  The terrestrial planets formed
from a subsequent, second generation of planetesimals.  We place the known
giant planets of the four systems examined on their best-fit (co-planar)
orbits and assign them their minimum masses (see Table 1).  To accurately
model terrestrial accretion in the system we need a complete understanding of
the planetary system, including the location of all mean motion and secular
resonances, which are determined by the true masses and relative inclinations
of all planets in the system.  Due to observational limitations and
incompleteness, we do not know $sin \, i$ or mutual inclinations of the
planets, a common limitation of many studies of extra-solar planets.  We
proceed under the assumption that the giant planets' orbits are co-planar and
that they are the only planets in the system.  If $sin\,i < 1$, the giant
planets' masses are larger than assumed here, and the stable zones from Papers
I and II would likely become narrower.  The discovery of additional planets
could potentially eliminate the stable zones completely.  But the goal of
these papers is to use dynamics to find additional planets; indeed, the stable
zones from Papers I and II are proposed as the most likely locations for
additional planets in these systems.

We assume a surface density profile of solids that decreases with
heliocentric distance as $r^{-3/2}$, i.e. surface density
$\Sigma$=$\Sigma_1\,r^{-3/2}$.  We first perform a set of simulations
assuming a substantial amount of mass remains in the terrestrial zone
after migration.  In these runs, the surface density is normalized at
1 AU to a value of $\Sigma_1$=10 $g\,cm^{-2}$.  This value is roughly
50\% larger than that for the minimum-mass solar nebula model
(Hayashi, 1981), and is reasonable under the assumption that the
surface density of solid material in the protoplanetary disk
correlates with stellar metallicity.  Indeed, three of the four known
host stars (all but HD 37124) have higher metallicities than the
Sun,\footnote{Data from http://www.exoplanets.org} and all four have
giant planets more massive than Jupiter.  For systems in which planets
accreted in the first set of simulations, we construct disks for a
second set of runs.  In this set, we assume that giant planet
migration severely depleted the planetesimal disk.  The remaining
surface density is normalized to 1.5 $g\,cm^{-2}$ at 1 AU, roughly one
quarter of the density of the MMSN model.  For late or slow giant
planet migration, Armitage (2003) does predict a drop of 1-2 orders of
magnitude in the surface density of planetesimals.  However, the mass
of planets that will form via accretion scales roughly linearly with
the surface density (Wetherill 1996; Kokubo \etal 2006).  Thus, we can
extrapolate from our simulations with $\Sigma_1 = 1.5 g\,cm^{-2}$ to
even more depleted disks.  We refer to our two sets of simulation as
representative of ``early migration'' (with a substantial planetesimal
disk) and ``late migration'' (with a depleted disk) scenarios.

\begin{figure*}
\centerline{
\plotone{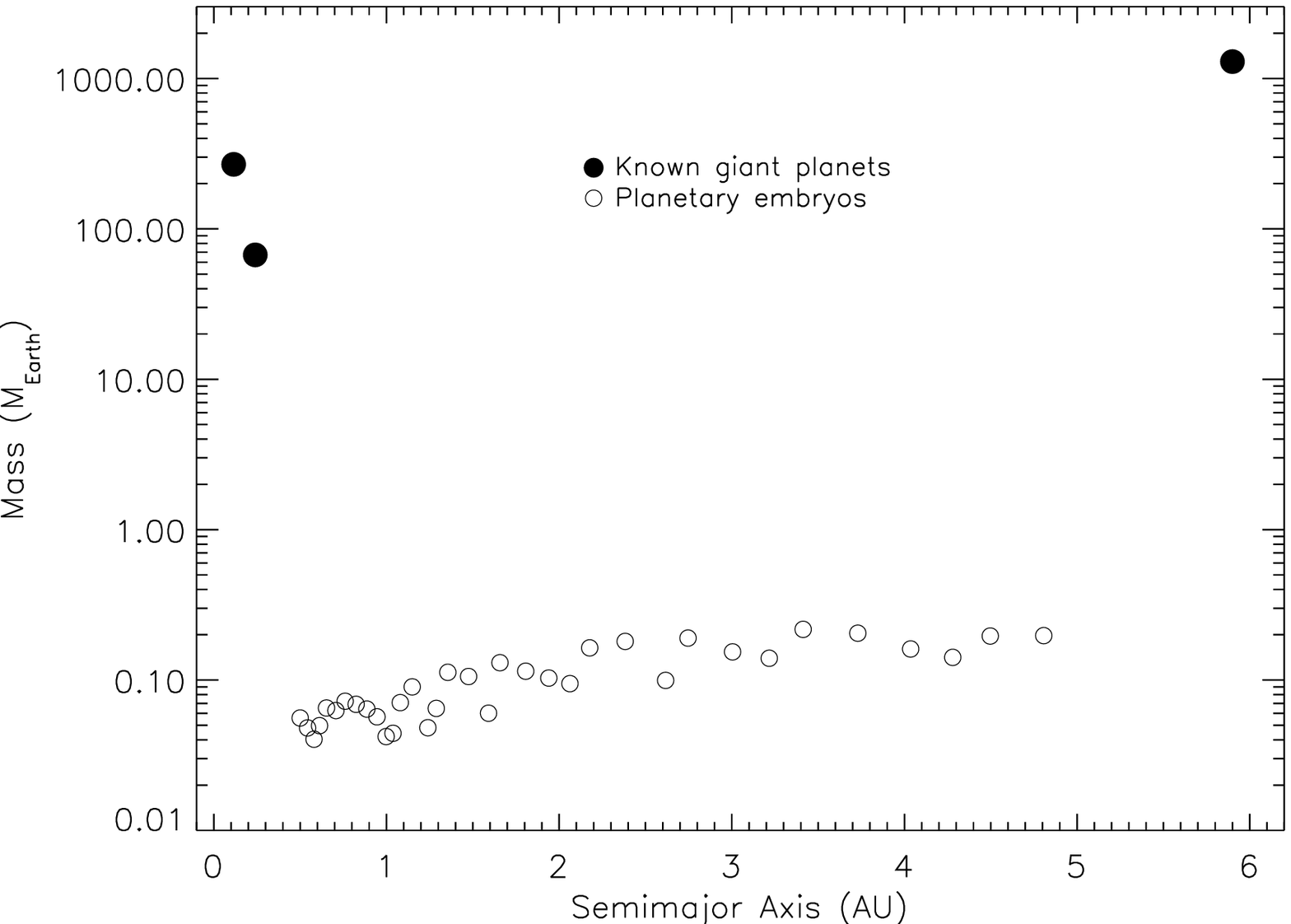}
}
\caption{Initial conditions for a simulation of 55 Cnc, assuming a surface
density of 10 $g\,cm^{-2}$ at 1 AU which scales with heliocentric
distance as $r^{-3/2}$. }
\label{fig:55i}
\end{figure*}

We assume that oligarchic growth has taken place in these systems, following
the formation of a second generation of planetesimals, as described in
Armitage (2003).  We place planetary embryos between the giant planets
following the method of Raymond, Quinn \& Lunine (2004, 2005a, 2006).  We
assume that embryos form with a typical spacing of $\Delta$ mutual Hill radii,
where $\Delta$=5-10 (as in, e.g., Kokubo \& Ida 2000).  The mass of embryos
$M$ scales as $M \propto r^{3/4} \Delta^{3/2} \Sigma_1^{3/2}$.  Thus, embryo
masses are larger farther from the star.  Figure~\ref{fig:55i} shows a set of
early migration initial conditions for a simulation of 55 Cnc.

The number of embryos we include depends on the separation of the giant
planets.  For the early migration simulations, we include 12-15 embryos in
HD 37124 of total mass $\sim$ 0.5 Earth masses between 0.8 and 1.2 AU; HD 38529:
25-29 embryos of total mass 2.7 $\mearth$ between 0.2 and 3.2 AU; HD 74156:
45-50 embryos of total mass 3.2 $\mearth$ between 0.6 and 3.6 AU; 55 Cnc: 34-38
embryos totaling 3.7 $\mearth$ between 0.5 and 4.9 AU.  Embryos are given
small initial eccentricities ($e \leq 0.02$) and inclinations ($i \leq
0.1^{\circ}$).  Note that some embryos are therefore initially placed outside
the stable regions from Papers I and II.  These embryos are quickly removed
from the system (usually via dynamical ejection); however, they allow us to
include populate the stable regions to their very edges.  The ejected embryos
do not affect the simulation outcome; indeed, it is not likley that embryos
could form in the harsh dynamical environment outside the stable zones.

We perform 10-12 early migration simulations for each planetary system to test
the likelihood of terrestrial planet formation.  In systems which formed
terrestrial planets (55 Cnc and HD 38529) we generate lower-mass disks of
embryos assuming a late migration scenario and run two additional simulations.
For 55 Cnc, we include 62-64 embryos between 0.5 and 5 AU totalling 1.1
$\mearth$ and for HD 38529 we include 63-66 embryos bewteen 0.2 and 2 AU
totalling 0.47 $\mearth$.  For the systems in which planets did not accrete
(HD 37124 and HD 74156), we run two additional simulations starting from a disk
of smaller bodies to see if any accretion is possible.

\subsection{Numerical Method}

We integrate the systems for 100-200 Myr using the hybrid integrator in
Mercury (Chambers 1999).  We chose a timestep for each system such that the
innermost initial orbit is sampled at least 20 times per orbit.  Collisions
conserve linear momentum and mass, and do not take collisional fragmentation
into account, but we discuss the distribution of impact angles and velocities
and the likelihood of accretional collisions.  Energy is conserved to better
than 1 part in 10$^4$.  Each simulation took 3-10 days to run on a desktop PC.

\section{Results}

We performed 10-12 simulations of terrestrial planet formation in each
of the four planetary systems listed in Table 1 assuming an early
giant planet migration scenario.  For systems in which accretion
proceeded we performed two additional simulations assuming late giant
planet migration.  For systems in which little accretion occurred, we
performed two additional higher-resolution simulations to test if any
accretion could happen.  

Here we summarize the results for each system.  An additional, $\sim$
Neptune-mass planet, 55 Cnc e, was discovered in this system in late
2004 (McArthur \etal, 2004).  It is not included in these simulations.
Because of its small mass, its large dynamical separation from the
terrestrial zone, and the small mass included in embryos, we do not
expect its presence to affect these results.

\subsection{55 Cancri}

\subsubsection{Early Migration}

\begin{figure*}
\centerline{
\plotone{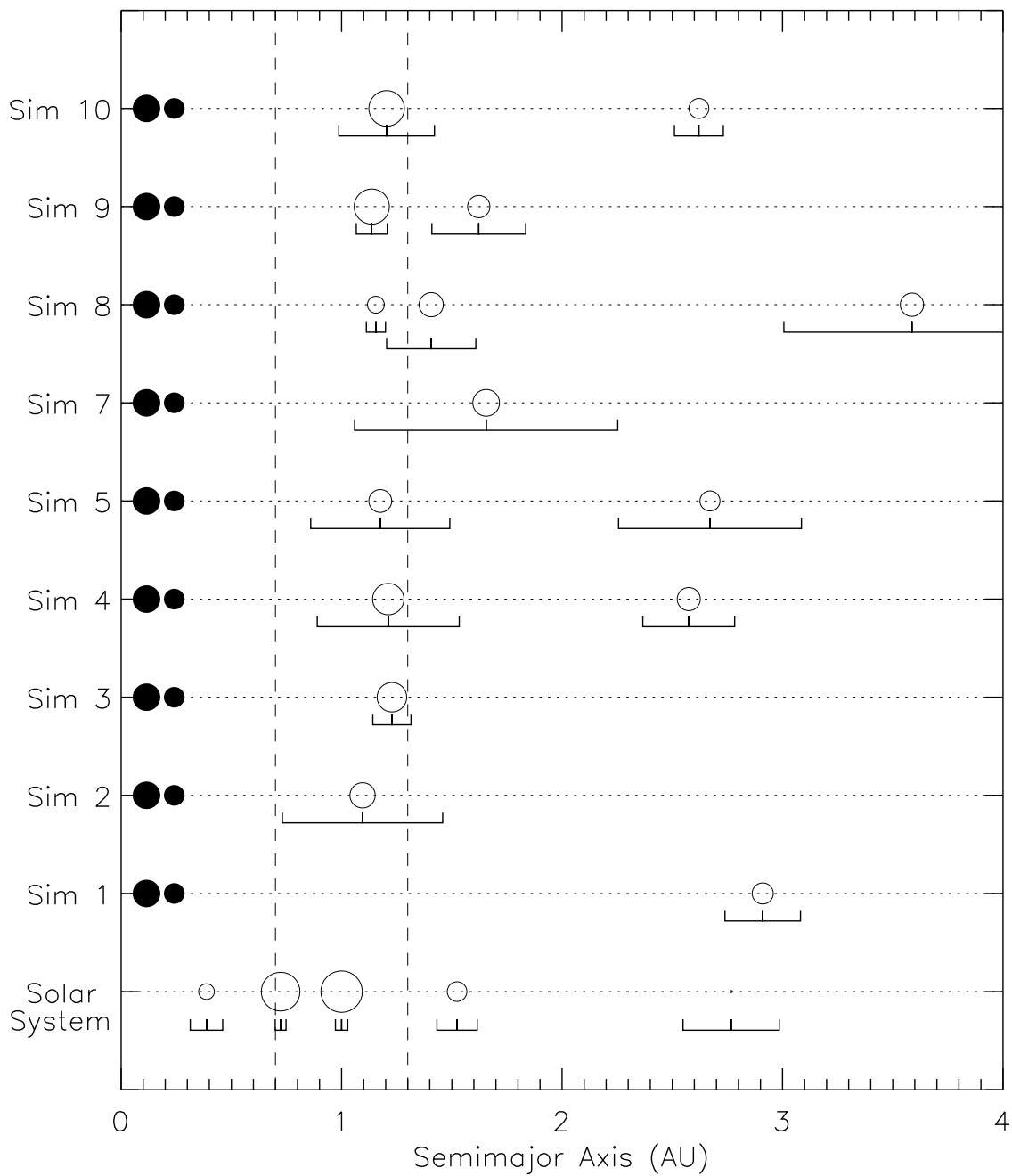}
}
\caption{Final configurations of all 9 simulations in 55 Cancri which formed
terrestrial planets.  The size of each circle is proportional to its
mass$^{(1/3)}$, and the underlying lines are bounded by each body's
perihelion and aphelion.  The dark solid circles indicate the
positions of the two inner known gas giant planets in the system,
whose sizes are not on the same scale as the terrestrial bodies (but
note that the inner one is more massive).  The Solar System (including
Ceres, the largest asteroid) is included for reference, with 3 Myr
averaged orbital values from Quinn \etal (1993).  The dashed vertical
lines represent the boundaries of the habitable zone in the system, as
quoted in Menou \& Tabachnik (2003).  The terrestrial planets which
formed in the habitable zones in simulations 5, 9 and 10 have accreted
material from past 2 AU, and have water mass fractions between
2$\times 10^{-4}$ and $10^{-3}$.}
\label{fig:55all}
\end{figure*}

Paper I revealed a wide stable region in 55 Cnc between 0.7 AU $ < a <$ 3.2 AU
with $e <$ 0.2.  We performed ten ``early migration'' terrestrial planet
formation simulations, nine of which formed 1-3 terrestrial planets with 1.1
AU $ < a <$ 3.6 AU, $e \leq$ 0.36, and masses up to 0.63 $\mearth$.
Figure~\ref{fig:55all} shows the final configurations of these nine
simulations, with the Solar System included for comparison.  Six simulations
finished with two or three remaining terrestrial bodies, including two systems
in which two planets had each accreted at least one embryo.  In two of the
nine cases the only embryos which remained had not accreted any other bodies.
In one case (simulation 2), the system took 170 Myr to reach a stable
configuration with no overlapping orbits.

The impact velocities for our early migration simulations of 55 Cancri are
similar to those seem by Agnor, Canup \& Levison (1999) in accretion
simulations including Jupiter and Saturn.  By examining the impact angles and
velocities, we estimate that 55-60\% of the impacts were accretional (i.e. the
final aggregate mass is larger than either impactor), following the results of
Agnor \& Asphaug (2004).  This is comparable to the 55\% found by Agnor \&
Asphaug (2004) for previous terrestrial accretion simulations that formed
planets similar to Venus, Earth and Mars.  We therefore consider the final
masses of our planets formed in 55 Cancri to be realistic.  Despite the fact
that we form terrestrial planets of significant mass ($>$ 0.5 $\mearth$), the
fraction of the total embryo mass which remains at the end of the simulations
is quite small, ranging from 4 to 20\% of the initial terrestrial mass.

\begin{figure*}
\epsscale{0.6}
\centerline{
\plotone{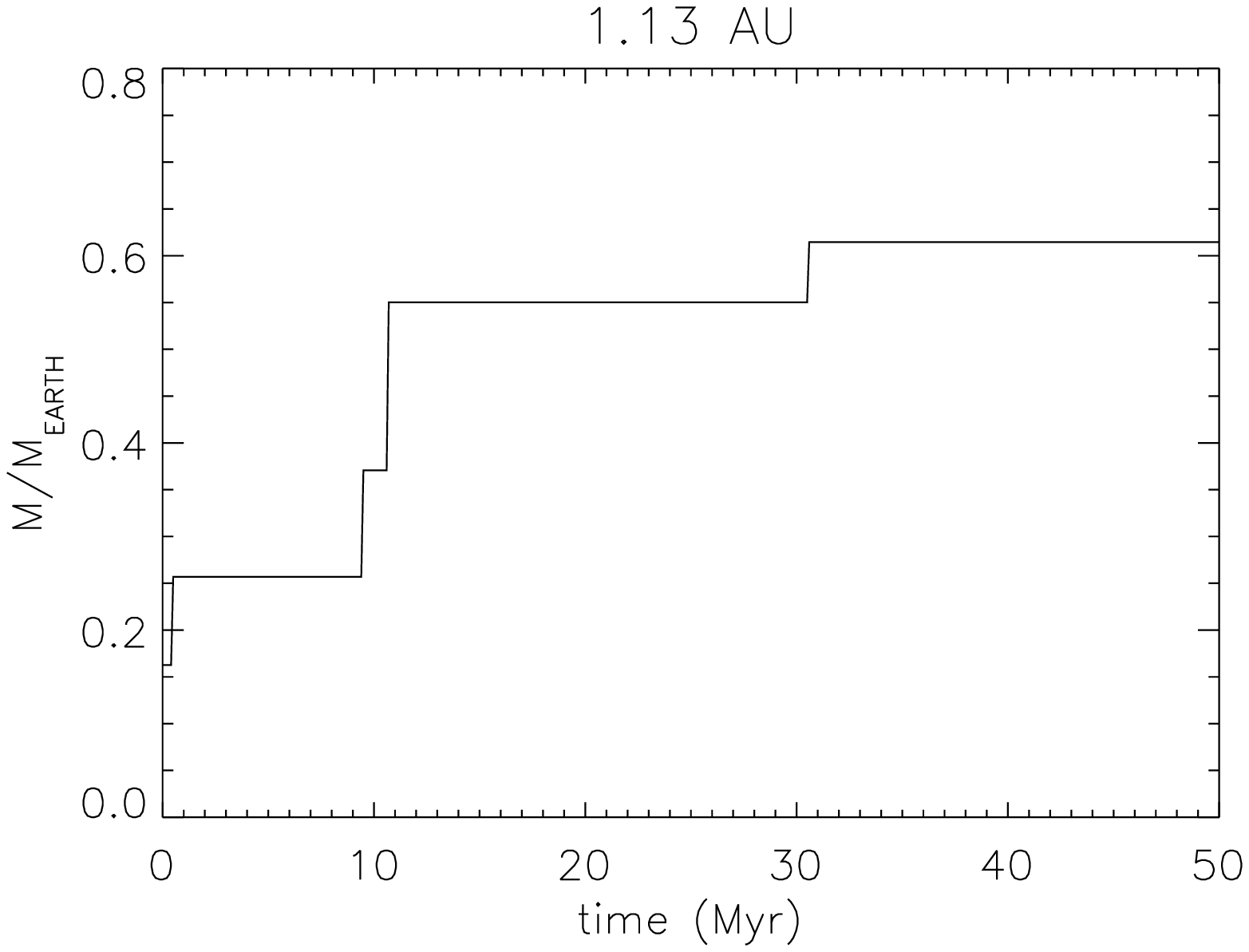}
}
\centerline{
\plotone{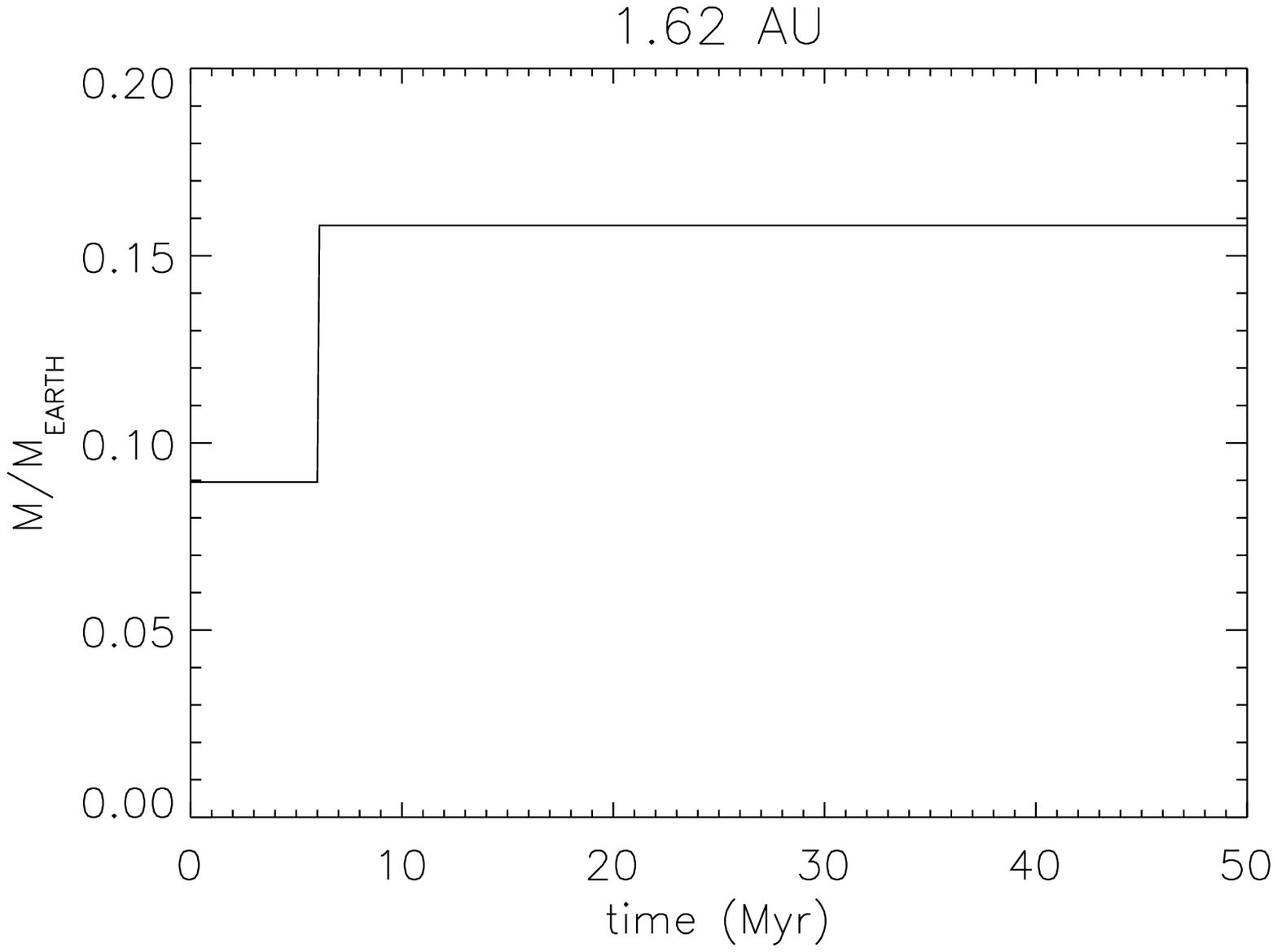}
}
\caption{The masses of two surviving terrestrial planets in 55 Cancri as a
function of time for the two surviving bodies in simulation 9 (see
Fig.~\ref{fig:55all}), labeled by their final semimajor axes. }
\label{fig:55form}
\end{figure*}

Figure~\ref{fig:55i} shows the initial configurations of a system (simulation
9 from Fig.~\ref{fig:55all}) which formed two terrestrial planets, with masses
of 0.61 $\mearth$ and 0.16 $\mearth$.  Figure~\ref{fig:55form} shows the
masses of these two planets as a function of time.  The larger undergoes four
accretion events, reaching its final mass at 30 Myr.  The smaller survivor is
an aggregation of only two embryos, which formed interior to the larger
survivor.  A close encounter at 70 Myr between the two final planets and
another embryo caused the two planets to roughly swap positions and the embryo
to be ejected from the system.  The planets' final orbital elements are ($a$,
$e$) = (1.14 AU, 0.06) and (1.62 AU, 0.13) for the larger and smaller planet,
respectively.


In our Solar system, chondritic material is seen to contain hydrated minerals
past 2-2.5 AU, with water mass fractions of up to $>$10\% for some
carbonaceous chondrites (Abe \etal 2000; see Fig. 2 from Raymond \etal, 2004).
The mass of the star $\rho$ Cancri is 0.95 M$_{\odot}$, so the temperature
distribution in its circumstellar disk should be similar to the solar nebula.
The temperature during the formation of the post-migration generation of
planetesimals would imprint a corresponding distribution of water content.  If
we apply the initial water distribution of Raymond \etal (2004) such that
embryos interior to 2 AU are dry, embryos between 2-2.5 AU contain 0.1\%
water, and those exterior to 2.5 AU contain 5\% water, then we can calculate
the water content of the terrestrial planets we have formed in 55 Cnc.  Of the
seven terrestrial planets whose semimajor axes lie in the habitable zone (see
Fig.~\ref{fig:55form}), three (simulations 5, 9, and 10) have accreted
material from past 2 AU and have water mass fractions between 2$\times
10^{-4}$ and $10^{-3}$ (the Earth's is roughly $10^{-3}$).  In particular,
note that the inner planet in simulation 9, shown in Figs.~\ref{fig:55all}
and~\ref{fig:55form}, accreted two embryos from past 2 AU and has a water mass
fraction of $7 \times 10^{-4}$.  This is the best candidate habitable planet
formed in our ten early migration simulations, as its mass is significant (0.6
$\mearth$), its eccentricity is small enough that its orbit stays inside the
habitable zone, and its water content is substantial.

\subsubsection{Late Migration}

We performed two additional late migration simulations in 55 Cnc
starting from a depleted disk.  Each simulation finished with 5-9
remaining bodies in the stable zone from Paper II.  The largest
surviving planet in either simulation accreted 9 other embryos but
reached only 0.13 $\mearth$.  If the disk were depleted by another
order of magnitude, then the most massive terrestrial bodies would
only be roughly a lunar mass.  Such planets are too small to have
long-lasting plate tectonics (Williams, Kasting \& Wade 1997) and are
unlikely to be habitable.\footnote{The mass cutoff for a planet to
have tectonic activity for 4-5 Gyr is though to lie roughly 0.3
$\mearth$, depending on the planetary density (Williams \etal 1997;
Raymond, Scalo \& Meadows 2006).}  In addition, this planet did not
accrete any water-rich bodies from past 2 AU and was dry.  Low-mass
disks do tend to form drier planets than more massive disks, because
the dynamical self-stirring of the disk is not efficient (Raymond
\etal 2004; Raymond, Scalo \& Meadows 2006).  However, an additional
source of gravitational stirring is present in the form of three giant
planets, but (free) eccentricities induced by the giant planets are
only $\sim$0.05-0.1 in the stable zones (Fig. 3 from Paper II), enough
to cause radial mixing on only a small scale.  The fact that several
``early migration'' planets did accrete water-rich embryos leads us to
the conclusion that, even in this complicated dynamical environment,
the self-gravity of the disk of embryos is important.  Embryos in
low-mass disks do not self-scatter enough to cause significant radial
mixing and water delivery.

\subsection{HD 38529}

\subsubsection{Early Migration}

\begin{figure*}
\epsscale{1.0}
\centerline{
\plotone{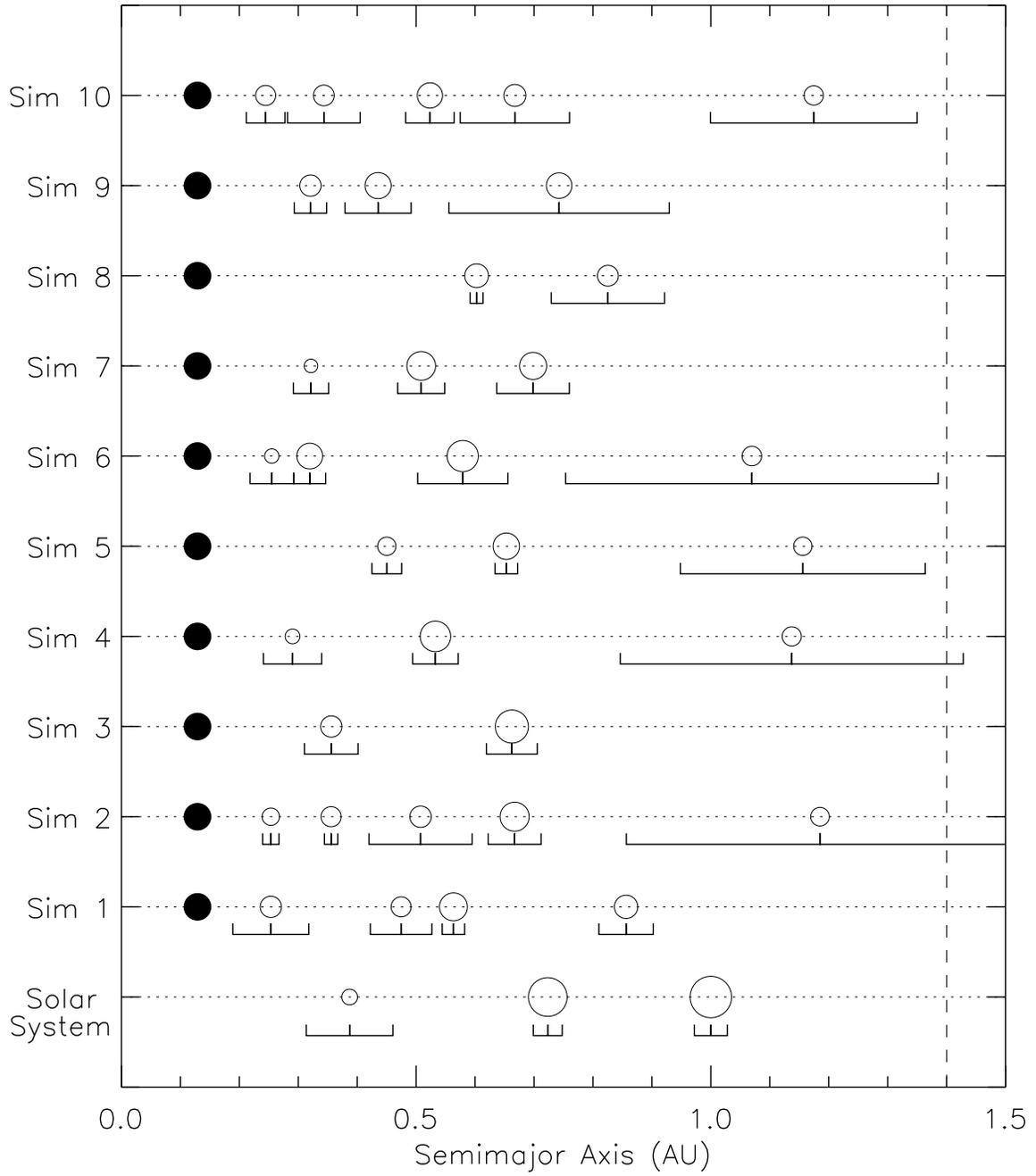}
}
\caption{Final configurations of all 10 simulations of HD 38529, formatted as
in Fig.~\ref{fig:55all}.  The solid circle represents the position of
the inner known giant planet.  The dashed vertical line at 1.4 AU
represents the inner edge of the habitable zone in this system, whose
host star's mass is 1.39 solar masses.}
\label{fig:38all}
\end{figure*}

Papers I and II found a large region in HD 38529 that was stable for
both massless and Saturn-mass particles.  This region contains two
dynamical distinct zones: from 0.3 AU to 0.5 AU planets are stable for
$e <$ 0.15, and from 0.5 to 0.8 AU they are stable for $e<0.3$.
Figure~\ref{fig:38all} summarizes the results of the ten early
migration simulations of HD 38529.  In each of the ten simulations,
2-5 terrestrial bodies remain after 200 Myr, averaging 3.4 surviving
bodies per simulation, with 0.04 $\mearth \leq M \leq$ 0.5 $\mearth$,
0.24 AU $< a <$ 1.2 AU, and $e \leq$ 0.3.  Most of the survivors have
accreted other bodies, and roughly one planet per system was massive
enough to be tectonically active ($>$0.2-0.3 $\mearth$; Williams \etal
1997, Raymond, Scalo \& Meadows 2006).  However, the final systems
look more like asteroid belts than planetary systems.  As discussed
below, this may be due to the development of chains of apsidal
libration which prevent bodies from undergoing close encounters.

In order for a relatively massive terrestrial planet to grow, it needs a
substantial feeding zone from which to accrete embryos.  A planet with a
higher eccentricity encounters embryos at a range of heliocentric distances,
and has the chance to accrete more embryos than a lower eccentricity planet
(e.g. Levison \& Agnor 2003).  However, the stable zones in HD 38529 extend
only to moderate eccentricities ($e<$0.15).  So, planets with higher
eccentricities and wider feeding zones are actually unstable.  Only planets at
relatively low eccentricities with smaller feeding zones can form in the
system.  

The impact velocities in HD 38529 are higher than for 55 Cancri.  This
is simply because the stable zone in HD 38529 extends to higher
eccentricities than in 55 Cnc.  Thus, embryos may interact and collide
with higher eccentricities and therefore at higher impact speeds.
Indeed, we estimate that only $\sim$ 30\% of collisions in HD 38529
are accretional, following the results of Agnor \& Asphaug (2004).  We
have not accounted for the expected debris and smaller bodies, which
may reduce the eccentricities of larger bodies via dynamical friction
(e.g., Goldreich, Lithwick, \& Sari, 2004).  We consider the masses of
the bodies we form in HD 38529 to be upper limits.  The total number
of surviving bodies may be a lower limit, reinforcing the possibility
that it is more likely for an asteroid belt to exist in HD 38529 than
terrestrial planets of significant size.

Simulation 10 from Fig.~\ref{fig:38all} is an interesting case, in which 5
terrestrial bodies remain, three of which have accreted at least one other
embryo.  The surviving bodies have semimajor axes of 0.24 AU, 0.34 AU, 0.52
AU, 0.67 AU, and 1.17 AU, and the known, giant planets have final semi-major
axes of 0.13 AU and 3.65 AU. At first glance this appears to be a chain of
mean-motion resonances amongst the inner planet and 4 innermost survivors with
period ratios of 5:2, 5:3, 2:1, and 10:7, respectively. However an inspection
of the resonance angles (see e.g.\ Murray and Dermott 1999, $\S$8.2) shows
that the relevant resonance angles are circulating, not librating.  This
configuration is therefore either a coincidence (note that the ratios are not
exactly in resonance), or these bodies are in a more complicated multi-body
resonance such as the three-body resonance between Jupiter, Saturn, and Uranus
in our solar system (Murray \& Holman 1999).

\begin{figure*}
\centerline{
\plotone{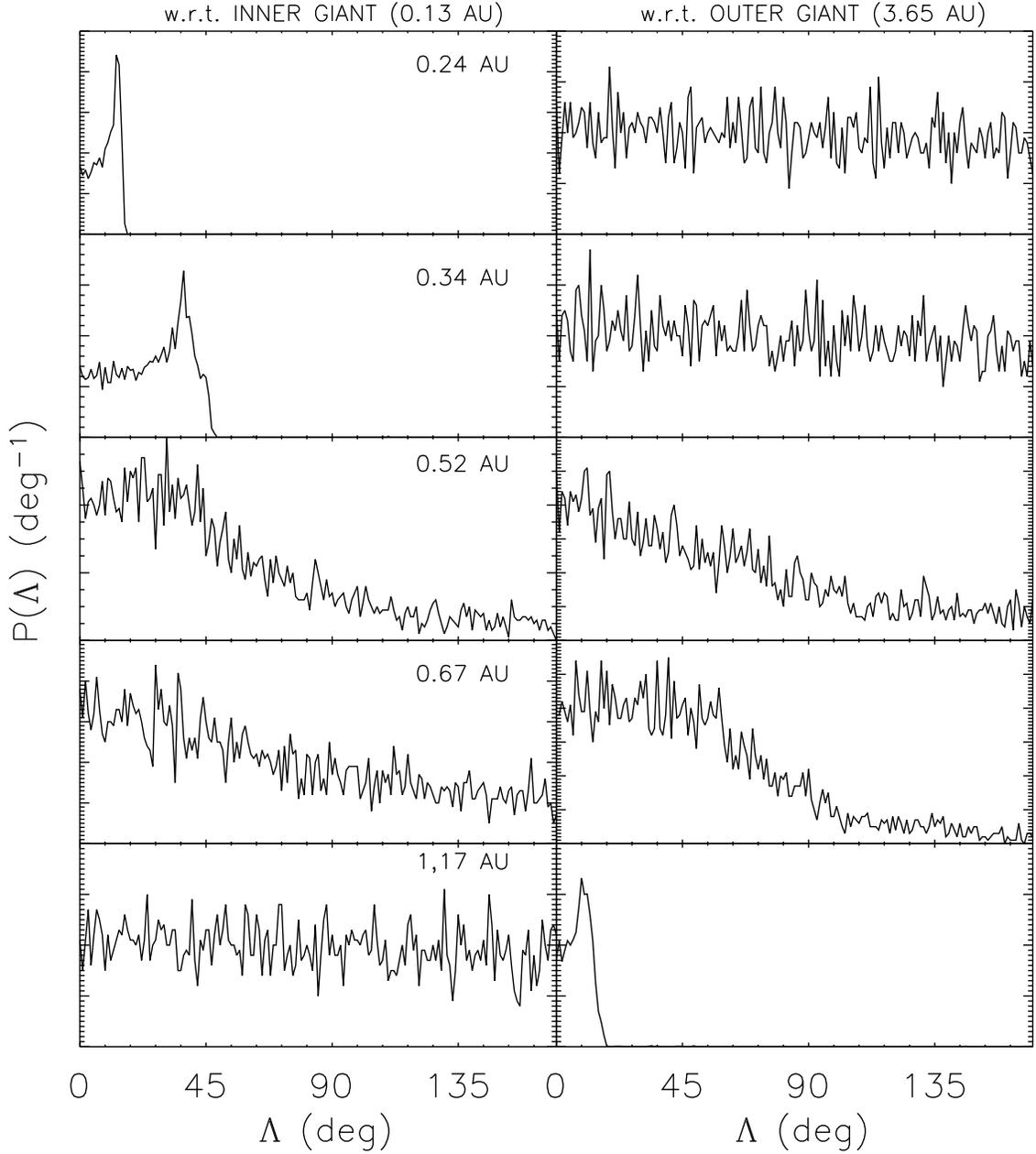}
}
\caption{The distribution function of the alignment of the longitudes
of pericenter for all surviving terrestrial bodies from simulation 10 of
HD 38529, formatted as in Fig.~\ref{fig:37a} (bottom panel).  Each row
corresponds to a given object: the left column shows the alignment with
respect to the inner giant planet at 0.12 AU, and the right column with
respect to the outer giant planet at 3.65 AU.  The inner bodies are in apsidal
libration with the inner giant planet, and the outer body with the outer
giant.}
\label{fig:secres10}
\end{figure*}

Apsidal libration, which occurs when two consecutive bodies'
longitudes of periastron oscillate about each other, pervades the
system. We search for these librations using the distribution function
of the relative orientation of planetary orbits.
Figure~\ref{fig:secres10} demonsrates libration among the surviving
bodies in simulation 10.  Each panel of Fig.~\ref{fig:secres10} shows
the distribution function $P(\Lambda)$, where $\Lambda$ is the
normalized difference between the longitude of pericenter of a
terrestrial body and that of a giant planet, as defined in Barnes \&
Quinn (2004).  Each row of the figure corresponds to a given surviving
terrestrial body in simulation 10, labeled by the body's semimajor
axis.  The left column represents the relative alignment of that
body's orbit with respect to the inner giant planet, and the right
column with respect to the outer giant planet.  A flat $P(\Lambda)$
curve indicates that the longitudes are circulating, while a sharp
peak indicates that a body's longitude of pericenter is librating
about that of the giant planet, and is therefore in apsidal libration.
It is clear from Fig.~\ref{fig:secres10} that the inner two
terrestrial bodies are in apsidal libration with the inner giant
planet, whose precession period is about 8 thousand years.  The
amplitude of libration of the innermost one is smaller, but both are
locked in this libration.  Similarly, the outermost terrestrial body
is in apsidal libration with the outer giant planet with a precession
period of 22 thousand years.  The two intermediate terrestrial bodies
are separated enough from both gas giants that there secular dynamics
are not dominated by either and their longitudes of periastron
circulate (as in, e.g., the bottom panel of Fig.5 from Paper II).

These librating configurations may be an additional factor in
explaining the low accretion rate in this system.  Orbits in apsidal
libration tend to avoid close encounters, because the orientation of
perihelion is the same for both orbits.  In simulation 10 we have
three such bodies in the inner system (giant planet plus two
terrestrial planets) and two in the outer system (outer giant and
terrestrial planets).  The place where collisions are most likely to
occur is at the juncture between these two regimes.  The two
terrestrial bodies in this juncture are indeed the most massive in the
simulation.  In almost every simulation the most massive terrestrial
planet lies in this middle region, between about 0.5 and 0.7 AU.
Interior to this astrocentric distance, bodies are likely to be in
apsidal with the inner planet (and each other), and outside with the
outer giant planet.  Of course, these librating configurations are
only induced after most accretion has happened in the system, and
objects are no longer on crossing orbits.

\subsubsection{Late Migration}

We performed two additional late migration simulations in HD 38529,
starting from a low-mass disk.  As in the early migration simulations,
3-4 bodies remained in each case.  The most massive of these accreted
11 other embryos but reached only 0.12 $\mearth$, below the likely
cutoff for habitability (Williams \etal 1997).  If the disk were
depleted by another order of magnitude, planets larger than the Moon
($\sim 0.013 \mearth$) could not form.

\subsection{HD 37124}

Paper I showed that the region in HD 37124 that is stable for massless
test particles lies roughly between 0.9 and 1.1 AU, with
eccentricities between 0 and 0.2.  Paper II showed that the likelihood
of a Saturn-mass planet surviving in this region of parameter space
for 100 Myr is roughly 50\%, with a maximum in survival rate between
0.9 and 1.2 AU at $e\sim$ 0.1.

\begin{figure*}
\centerline{
\epsscale{0.6}
\plotone{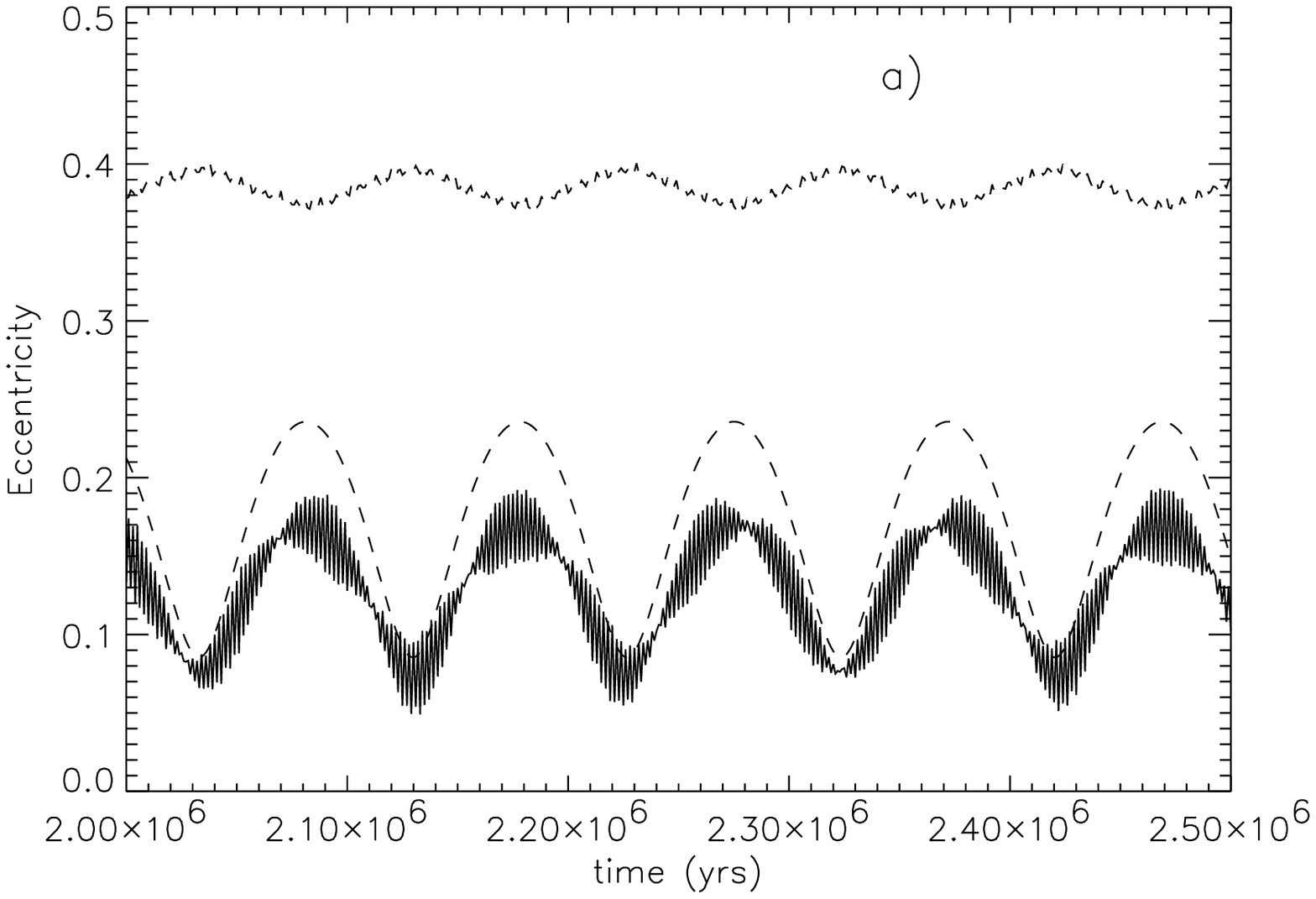}
}
\centerline{
\plotone{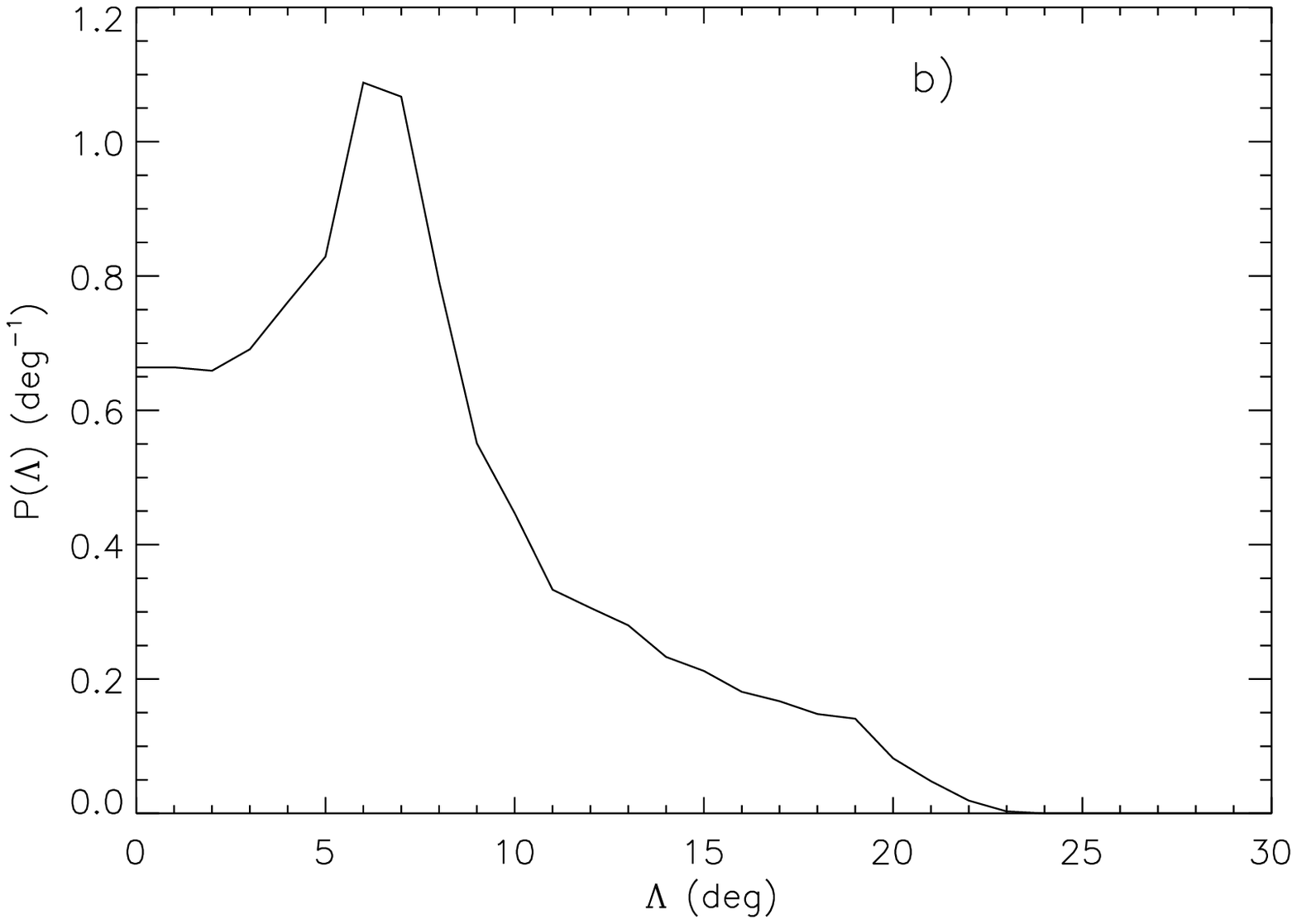}
}
\caption{\underline {Top}: Eccentricity evolution of the two giant
planets (dashed lines) and one remaining planetary embryo (solid line)
with a semimajor axis of 0.95 AU in a simulation of HD 37124.  The
period of the giant planets' eccentricity variations is 96 kyr.
\underline {Bottom}: The distribution function, P($\Lambda$), of the
same embryo as a function of $\Lambda$, the normalized difference between
their longitude of pericenter and that of the inner planet, as in
Fig.~\ref{fig:secres10}.  The embryo is in apsidal libration with the inner
giant planet.  The shape of the P($\Lambda$) curve, similar to that of a
harmonic oscillator, indicates that the longitude of pericenter of the
embryo's orbit is librating with respect to the giant planet's orbit.}
\label{fig:37a}
\end{figure*}

In twelve early migration simulations in HD 37124, almost all embryos
were ejected, and no terrestrial planets formed.  However, in seven
cases, a solitary embryo survived to the end of the simulation.  All
such embryos had final semimajor axes 0.9 AU $ < a <$ 1.1 AU and
eccentricities 0.1 $< e <$ 0.22.  Additionally, in each case the
surviving embryo appears to be in apsidal libration with the inner
giant planet.  Figure~\ref{fig:37a}a (top panel) shows the
eccentricity evolution of a remaining embryo with $a$ = 0.95 AU, which
is following the eccentricity of the inner planet closely, with
superposed higher order fluctuations.  The periods of the eccentricity
variations of the giant planets are both 96 kyr.  The apsidal nodes of
the giant planets are librating with an amplitude of 31$^{\circ}$.
The remaining embryo is in apsidal libration with both giant planets,
but more strongly with the inner planet.  Figure~\ref{fig:37a}b shows
the distribution function of $\Lambda$, the difference between the
longitude of pericenter of the embryo and that of the inner planet, as
in Fig.~\ref{fig:secres10}.  The sharp peak at $\Lambda \simeq$
6$^{\circ}$ indicates that the embryo's longitude of pericenter is
librating about that of the inner giant planet.  This is not
unexpected given the results of Paper II, in which we showed that
apsidal libration in HD 37124 plays an important role in the stability
of Saturn-mass planets.

We performed two additional higher resolution simulations in this system to
verify these results.  We placed 100 embryos of mass 0.005 $\mearth$ between
0.9 and 1.1 AU and integrated the system for 100 Myr.  In one case, no embryos
survived to the end of the integration.  In the other, one embryo survived,
again in apsidal libration with the inner giant planet.  This embryo accreted
one other embryo and had a final mass of 0.01 $\mearth$, roughly that of the
Moon.  This indicates that {\it in situ} formation of terrestrial planets in
this system is unlikely.

\subsection{HD 74156} 

Paper I found a very limited stable region for massless test particles in
HD 74156.  In Paper II, we showed that roughly 40\% of Saturn-mass planets are
stable for 0.5 AU $ < a <$ 1.5 AU, with a broad peak in survival rate at 0.9
AU $ < a <$ 1.4 AU.

As in HD 37124, no terrestrial planets formed in any of the ten simulations of
this system, as the vast majority of embryos were ejected.  In six cases, one
embryo survived until the end of the simulation, with final orbital elements
0.8 AU $ < a <$ 1.5 AU and 0.09 $< e <$ 0.17, consistent with Paper II.

We again performed two additional simulations with 100 0.02 $\mearth$ embryos
spaced randomly between 0.5 and 1.5 AU.  In one case one embryo remained after
100 Myr, and in the other, two were left.  In neither case had any surviving
terrestrial body accreted another embryo.  In each case the final orbital
parameters were similar to the survivors in the early migration simulations,
indicating that in situ formation of terrestrial planets in HD 74156 is
unlikely.

\section{Discussion \& Conclusions}

Our simulations show that certain systems of giant planets are conducive to
forming terrestrial planets, others are likely to contain belts of debris or
asteroids, and some are not likely to contain any rocky bodies.  Our systems
of giant planets are drawn from observations, although we have made the
important assumption that these systems are complete and have well-determined
orbits.  We have used slightly outdated orbital parameters in order to remain
consistent with previous work (specifically papers I and II).  If, for
example, new values of $sin \, i$ were determined for 55 Cnc and HD 38529, it
would narrow the stable zones for additional planets and affect the region in
which terrestrial planets could form. 

Our ``early migration'' simulations follow the reasoning of Raymond \etal
(2005a), who argue that if giant planets form and migrate in less than roughly
1 Myr, then terrestrial planets may form via accretion in the standard way
from a second generation of planetesimals (Armitage, 2003).  Our simulations
therefore start with the gas giant planets already present, although their
formation mechanism is unknown, be it core accretion (e.g. Rice \& Armitage
2003) or gravitational collapse (e.g. Mayer \etal 2002).  In either scenario,
it is likely that the giant planets formed farther out in the disk and
migrated inward via interactions with the gaseous disk (e.g. Lin \etal 1996).

If giant planet migration occurred late in the evolution of
protoplanetary disks, then the planetesimal disk would be severely
depleted (Armitage 2003), and it is unlikely that habitable planets
could form in any of the systems studied here.  The largest mass of
planets that form in our ``late migration'' simulations is below the
predicted lower limits of 0.2-0.3 $\mearth$ for a 'tectonic' habitable
planet (Williams \etal 1997; Raymond, Scalo \& Meadows 2006).  Late or
slow migration could potentially deplete the disk by an order of
magnitude more than we have simulated.  Since planet mass scales
roughly linearly with surface density (Wetherill 1996; Kokubo \etal
2006), these very low mass disks aren't capable of forming planets
much more massive than the Moon.

Our simulations of 55 Cancri suggest that a potentially habitable planet could
form {\it in situ}.  Such a planet would have a small enough eccentricity to
remain in the habitable zone throughout its orbit, substantial mass and water
content.  However, as shown in Paper II, a Saturn-mass planet could exist on a
stable orbit in the habitable zone of 55 Cnc.  Such a planet may preclude the
existence of habitable planet, although there remains the possibility of a
habitable satellite of the giant planet (Williams \etal 1997).

The systems of terresrtial planets formed in 55 Cnc (and to a lesser extent in
HD 38529) do not ressemble planets formed in preivous dynamical simulations
(e.g. Agnor \etal 1999; Chambers 2001; Raymond, Quinn \& Lunine 2004, 2005a,
2005b, 2006).  Indeed, previous simulations tend to include systems of giant
planets similar to Jupiter and Saturn, with relatively low eccentricities.
The large masses and higher eccentricities of planets in 55 Cnc and HD 38529
increase the perturbations felt by embryos, causing a much higher rate of
dynamical ejection than in previous simulations.  The zones in which accretion
can occur correspond roughly to the stable zones from Papers I and II, and are
much narrower than for systems with only an interior or exterior giant
planets.  Thus, planets that form in 55 Cnc and HD 38529 are significantly
smaller than in previous simulations for the same mass disk.  In addition, the
distribution of planet masses tends to peak near the center of the stable
regions from Papers I and II.  Strong perturbations mark the boundaries of the
stable regions, so embryos that stray from the edges are quickly ejected.

The case of HD 38529 is an interesting one.  Several terrestrial bodies survive
in our simulations of HD 38529, but do not accrete into large planets.  Despite
strong giant planet perturbations, growing terrestrial planets do not reach
high enough eccentricities to widen their feeding zones sufficiently to form
Earth-sized planets.  This is likely because the stable region from Papers I
and II extends only to eccentricities of 0.15 (0.3 in some areas).  Thus,
planets which reach these high eccentricities are ejected rather than
accreting into large terrestrial planets. We therefore speculate that a
well-populated asteroid belt may exist in HD 38529, potentially including
several Mars-sized planets but no Earth-sized planets.

Paper II found a wide zone ($0.3<a<0.8, e<0.15$) in HD 38529 stable for Saturn-
or even Jupiter-mass planets.  Observations have ruled out the existence of
such a massive planet, but not of a $\sim$Neptune-mass planet.  However, it
appears unlikely that accretion could form a large planet {\it in situ}.  How
do we reconcile this with the PPS hypothesis?  There certainly exists the
possibility that a massive planet co-migrated with the inner giant planet,
settling into the stable region (e.g., Thommes 2005).  Indeed, PPS predicts
that such a large contiguous stable region {\it must} contain a planet with
significant mass.

This is the third paper of the Predicting Planets series.  We have conducted
tests of dynamical stability and accretion in four systems of known
extra-solar giant planets -- HD 37124, HD 38529, HD 74156, and 55 Cancri.  We
have searched for regions in which the orbits of massless test particles are
stable for millions of years (Paper I), which may correspond to stable
Earth-mass planets.  We have self-consistently tested for regions which are
stable for Saturn-mass test particles (Paper II).  In this paper, we have
demonstrated that terrestrial planets may be able to form by accretion in one
of these systems, 55 Cancri.

This series of paper has laid the foundation, along with Barnes \&
Quinn (2004), for the PPS hypothesis. These papers have clarified the
conditions necessary for this hypothesis to be (dis)proven. We have
also seen that some systems which initially appeared to contain stable
regions, such as HD 168443, which was shown in Barnes \& Quinn (2004)
to lie far from instability, are in fact packed (Paper I; to be packed
to the point that test particles cannot survive for even $10^6$
years). We have also shown that two formation avenues exist for a PPS
system: simultaneous or episodic. Comparing Paper II to this work, we
see that the only possibility for HD 37124 and HD 74156 to satisfy the
PPS hypothesis is for an undetected giant planet to form coevally with
the known planets.  However in 55 Cnc and HD 38529, additional planets
could have formed and evolved together with the known planets, or
possibly formed \textit{after} a quick migratory epoch. In its
current, nascent state, the PPS hypothesis cannot distinguish between
these two formation events. If planets are found where we predict them
to be in the 55 Cnc or HD 38529 system, then nothing can be said about
their formation. Should they be found in HD 37124 or HD 74156, then
only coeval formation is possible.

It remains to be seen whether the predicted planets exist.  We do not predict
that stable regions will contain so much mass as to be borderline unstable.
Rather, we suggest that any region in a planetary system which can support a
massive planet will contain a planet.  Future observations of these
well-studied planetary systems will test the credence of the PPS hypothesis.

\section{Acknowledgements}
We thank the anonymous referee for helpful comments, Tom Quinn for useful
discussions, and Chance Reschke for his assistance in the completion of the
simulations presented in this paper. This work was funded by grants from the
NASA Astrobiology Institute, the NSF, and a NASA GSRP. These simulations were
performed on computers donated by the University of Washington Student
Technology Fund performed under CONDOR.\footnote{CONDOR is publicly available
at http://www.cs.wisc.edu/condor}

\scriptsize
\begin{deluxetable}{ccccccccc}
\tablewidth{0pt}
\tablecaption{Orbital Parameters of Selected Planetary Systems}
\renewcommand{\arraystretch}{.6}
\tablehead{
\\
\colhead{System} &  
\colhead{Planet} & 
\colhead{M ($M_J$)} &
\colhead{$a$ (AU)} &  
\colhead{$e$}&
\colhead{$\varpi (^\circ)$}&
\colhead{T (JD)} &
\colhead{$M_\star(M_\odot)$}\tablenotemark{1} &
\colhead{[Fe/H]}}
\startdata

HD 37124\tablenotemark{2} & b & 0.86 & 0.54 & 0.1 & 97.0 & 2451227 & 0.91 & -0.32\\
 & c & 1.01 & 2.95 & 0.4 & 265.0 & 2451828\\
\hline
HD 38529 & b & 0.78 & 0.129 & 0.29 & 87.7 & 2450005.8 & 1.39 & 0.313\\
 & c & 12.8 & 3.68 & 0.36 & 14.7 & 2450073.8\\
\hline
55 Cnc\tablenotemark{3} & b & 0.84 & 0.115 & 0.02 & 99.0 & 2450001.479 & 1.03 &
 0.29\\
 & c & 0.21 & 0.241 & 0.339 & 61.0 & 2450031.4\\
 & d & 4.05 & 5.9 & 0.16 & 201.0 & 2452785\\
\hline
HD 74156\tablenotemark{4} & b & 1.61 & 0.28 & 0.647 & 185.0 & 2451981.38 & 1.05
 & 0.13\\
 & c & 8.21 & 3.82 & 0.354 & 272.0 & 2451012.0\\
\enddata
\tablenotetext{1}{Stellar masses $M_\star$ and metallicities [Fe/H] are from
 the extra solar planets encyclopedia: www.obspm.fr/planets}
\tablenotetext{2}{Best fit values for HD 37124 from Butler \etal (2003).  The
 current best fit  for planet c is $a$ = 2.50 AU, $e$ = 0.69
 (http://www.exoplanets.org)} 
\tablenotetext{3}{Another planet, 55 Cnc e, was discovered in this
  system in late 2004 (McArthur \etal, 2004).}
\tablenotetext{4}{Best fit values for HD 74156 as of August 22, 2002.  The
 current best fit  for planet c is $a$ = 3.40 AU, $e$ = 0.58 (Naef \etal
 2003).} 
\end{deluxetable}

\end{document}